\documentclass[11pt]{article}
\usepackage{times}
\usepackage{epsf}
\usepackage{latexsym}
\usepackage{amsmath}
\usepackage{amsfonts}
\usepackage{euscript}
\usepackage{url}
\usepackage{graphicx}
\usepackage{subfigure}
\usepackage{enumitem}
\newcommand{\ignore}[1]{}

\begin{document}

\title{Toward Safe Integration of Legacy SCADA Systems in the Smart Grid}

\author{Aldar C-F. Chan\\
University of Hong Kong 
\and 
Jianying Zhou\\
Singapore University of Technology and Design
}

\maketitle

\begin{abstract}
A SCADA system is a distributed network of cyber-physical devices used for instrumentation and control of critical infrastructures such as an electric power grid.  With the emergence of the smart grid, SCADA systems are increasingly required to be connected to more open systems and security becomes crucial.  However, many of these SCADA systems have been deployed for decades and were initially not designed with security in mind.  In particular, the field devices in these systems are vulnerable to false command injection from an intruding or compromised device.  But implementing cryptographic defence on these old-generation devices is challenging due to their computation constraints.  As a key requirement, solutions to protect legacy SCADA systems have to be an add-on.  This paper discusses two add-on defence strategies for legacy SCADA systems --- the data diode and the detect-and-respond approach --- and compares their security guarantees and applicable scenarios.  A generic architectural framework is also proposed to implement the detect-and-respond strategy, with an instantiation to demonstrate its practicality.
\end{abstract}

%\begin{keyword}
%SCADA; power grid; cybersecurity; data diode; detect-and-respond
%\end{keyword}

\section{Introduction}
\label{sect:introduction}
A SCADA (Supervisory, Control And Data Acquisition) system is a cyber network of communicating devices used for instrumentation and control of a distributed infrastructure, such as an electric power grid, in order to manage the respective physical processes.  The SCADA system of a power grid can monitor and control various electric equipment along the power delivery path --- including various transformers, circuit breakers, protective relays and automatic re-closers --- while acquiring different measurements, which provide information about the loading conditions at different parts of the power grid and health conditions of transmission lines, etc.  Hence, correct operation of a SCADA system is critical to the reliability of any power distribution system \cite{ALW16,ALZR11}.

A successful malicious attack targeting at the SCADA system could potentially cause extensive power outage.  Such service disruption could even have cascading effects onto other critical infrastructures \cite{ALZR11}.  However, most of the SCADA systems currently in active use have been deployed for decades and very few of them were originally designed with security in mind \cite{MKWDSMK16,NPP17,WKM04}.  These legacy SCADA systems were assumed to operate in isolation at first, with obscurity and physical isolation being the main security strategy.  But nowadays, this design assumption is challenged not only by new security threats --- including demonstrated break-in’s of power substations \cite{T14}, discovered worms on PLCs (Programmable Logic Controllers) and other industrial control system platforms \cite{L11,NM18}, and the exposure of supposedly obstructed documents (such as SCADA system configuration and operation manuals) on the Internet \cite{MKWDSMK16} --- but also by the changing operational model which desires a higher level of integration with relatively open systems \cite{HKY11}, such as the AMI (Advanced Metering Infrastructure), in the envisioned smart grid for data sharing and orchestration of process control.

The increasing connectivity of these legacy SCADA systems with external smart grid systems, with comparatively more open access, could increase the attack surface for attackers to compromise devices inside the SCADA systems or send forged data or commands directly from outside.  Nonetheless, these legacy SCADA systems will take decades to be phased out from operation since they are so highly integrated with the power equipment they monitor \cite{CATO17,NPP17}.  It is thus essential to devise protection mechanisms to safeguard legacy SCADA systems as they are increasingly integrated with newer smart grid systems.

In particular, legacy SCADA systems are typically vulnerable to forged data or commands sent from intruding or compromised devices (i.e. insider attacks) \cite{ALW16,CATO17,HMW17,MKWDSMK16,NPP17,PKK13}.  For instance, a compromised SCADA RTU (Remote Terminal Unit) could launch a false command attack whereby an attacker impersonates the master station to send false control commands to other RTUs, say, to maliciously trip a particular circuit breaker.  On the other hand, false readings from compromised instrumentation devices could present a distorted picture of the system status to the master station, thus possibly triggering a false alarm and disruptive actions \cite{HMW17}.  Currently, most legacy SCADA systems could have little defense to this kind of attacks as no source authentication or command verification mechanism is in place \cite{ALW16,ACF16,CATO17,MKWDSMK16,NPP17,PKK13}.

Implementing cryptographic defence on the old-generation devices in these SCADA systems is impossible due to their resource constraints, and protection has to resort to the bump-in-the-wire approach \cite{WKM04}, which is generally regarded as very inefficient.  While authentication and authorization frameworks (such as IEC62351-8) --- which are increasingly applied for authentication and access policy enforcement in distributed smart grid and newer SCADA systems \cite{ALW16,PKK13} --- can avoid or minimize false command injection, it remains challenging to deploy them in some legacy SCADA systems and protocols due to the computational and communication overheads involved, as well as, the need of modifications at both the protocol and device level.  To avoid protocol modification, legacy-compliant message authentication \cite{CATO17} is proposed to embed authentication data as an additional payload of some specific SCADA protocols.  However, it requires modifications of the device software and is applicable only to a few, usually newer SCADA protocols and devices that are powerful enough to run common cryptographic primitives.  Generic strategies through enhanced incident management \cite{AANL11} or better process monitoring \cite{ALZR11} could improve SCADA system security but require a higher degree of human involvement and intervention.  It is fair to say truly add-on, non-intrusive protection for legacy SCADA systems and protocols remains challenging.

This paper studies and compares two different strategies to secure legacy SCADA systems for safe integration with smart grid systems.  Both approaches do not require any modification on the devices, protocols or communication channels, and are scalable for different numbers of devices.  The need of human intervention is also minimized.  The first approach, called the “data diode” approach \cite{FRCS19,OS10,YCKK16}, is commonly adopted in the industry and aims to preserve the isolation of a legacy SCADA system while facilitating outward information flow to newer smart grid systems.  The second approach aims to open up a legacy SCADA system for bidirectional information flow, with mechanisms in place to detect and identify false commands sent by intruding devices or compromised nodes.  This paper proposes a high-level, architectural framework for the second approach to implement a detect-and-respond strategy to neutralize the effects of false command attacks on a SCADA field network.  While this paper focuses on false command detection based on verification against an authenticated copy of each command received through a secure channel, the framework provides a general basis for incorporating other attack detection methods \cite{HMW17,LSKSI16,HM18} to implement a defence strategy.  This paper will also compare the applicability and usability of the two approaches.

The contributions of this paper is two-fold.  First, a generic architectural framework is proposed to implement the detect-and-respond strategy for the protection of legacy SCADA systems against false command attacks, with a view to opening up these SCADA systems for integration with relatively more open smart grid systems.  The proposed technique is non-intrusive and truly add-on without requiring modifications on existing devices and systems.  It is also scalable in the sense that only one defending device is needed per field network.  To demonstrate the practicality of the proposed framework, an instantiation of the detect-and-respond strategy on the Siemens Sinaut 8FW protocol \cite{S01} --- a common industrial protocol for legacy SCADA systems --- is presented with performance results.  Second, this paper presents a comprehensive review of the current landscape of research in legacy SCADA system protection.  It compares the data diode and detect-and-respond approaches, discussing their benefits and limitations, and provides practical guidelines for their application.

The rest of this paper is organized as follows.  In the next section, legacy SCADA systems and false command attacks against them are discussed.  The data diode strategy and the detect-and-respond strategy are presented in Section~\ref{sect:data-diode} and Section~\ref{sect:detect-and-respond} respectively.  Finally, Section~\ref{sect:conclusions} concludes with a comparison of the two strategies.

\section{False Command Attacks against Legacy SCADA Systems}
\label{sect:false_command}
\begin{figure}[htbp]
    \centering
    \includegraphics[width=14.5cm]{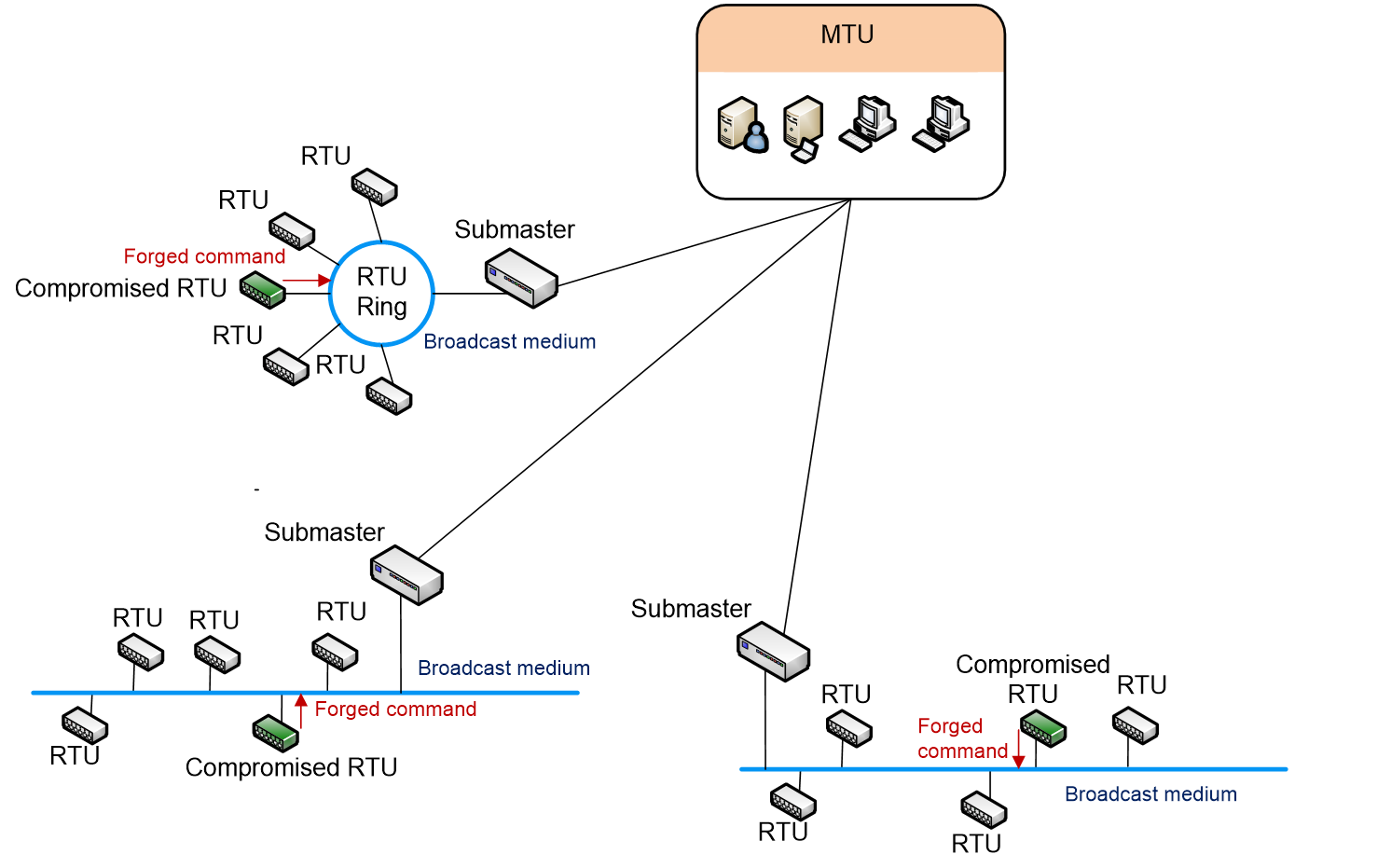}
    \vskip -0.3cm
    \caption{A typical legacy SCADA system and the possibility of false command attacks}
    \label{fig:scada_network}
\end{figure}

\begin{figure}[htbp]
    \centering
    \includegraphics[width=10cm]{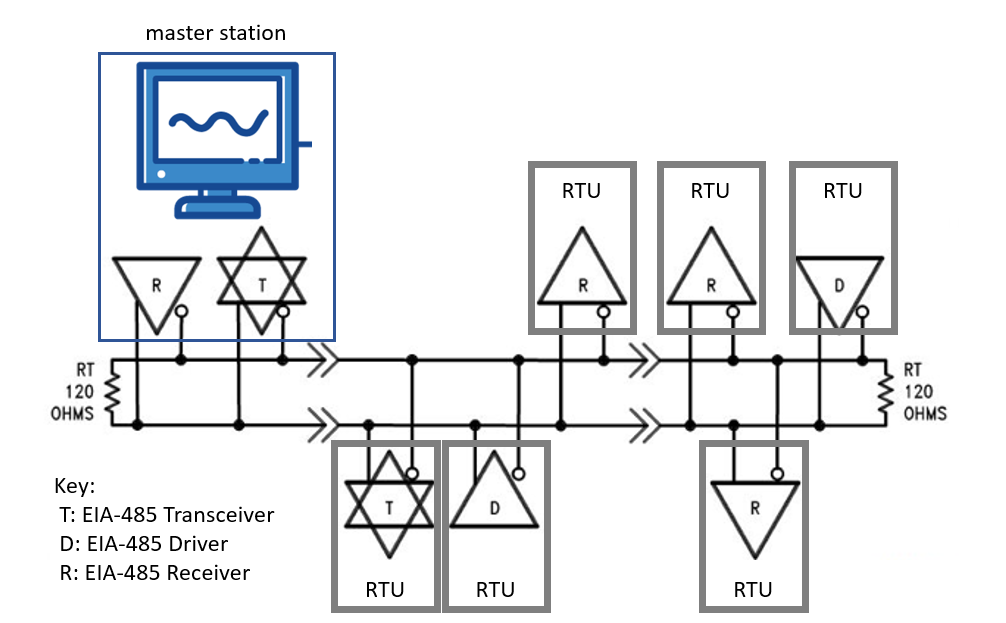}
    %\vskip -0.3cm
    \caption{EIA-485 serial ports connected in the party-line mode}
    \label{fig:eia485}
\end{figure}

A typical SCADA system, as depicted in Figure~\ref{fig:scada_network}, consists of PLCs and RTUs in the older versions and IEDs (Intelligent Electronic Devices) in the newer versions as the basic units for deployment in power substations or remote sites (for monitoring transmission lines) at different geographic locations.  Each of these field devices has interfaces to sensors and actuators used to monitor and control equipment in a power delivery system.  The field devices in a substation or remote site form a local area network, known as the field network, and are usually connected together over a broadcast medium to a sub-master or data concentrator, which in turn is connected to the MTU (Master Terminal Unit) or master station at the main control center through leased lines over a telecom operator’s network.  While various topologies are in use for connecting field devices with the sub-master, a broadcast channel --- such as EIA-485 serial ports in the party-line mode over a pilot cable (as shown in Figure~\ref{fig:eia485}) or DNP3 in the multi-drop or data concentrator mode\footnote{Earlier SCADA systems were designed based on proprietary protocols.  DNP3 is a standard adopted in newer SCADA systems for connecting RTUs and IEDs with an MTU.} --- is typically used in SCADA systems.

\vskip 1cm

Despite cost efficiency achieved in cable layout and maintenance, the use of broadcast channels also eases insider attacks from compromised devices.  For example, a compromised RTU can pose the MTU to issue a forged command to another RTU or PLC, causing the latter to behave abnormally.  Experiments have been repeatedly reported to demonstrate the possibility of using off-the-shelf protocol analyzer software to intercept control messages and inject false commands in some legacy SCADA systems \cite{HMW17}, with some showing the possibility to trip a circuit breaker maliciously.  In a few cases, even though the circuit breaker signaled the master station that it had opened, the master station did not respond as it had not instructed the circuit breaker to open.

This type of attacks was inconceivable in the original design of SCADA systems, and SCADA devices have little resource to defend against them.  Obscurity in system design, command formats and operation, combined with physical isolation, have long been the main and possibly the only strategy for securing legacy SCADA systems.  However, the effectiveness of obscurity may have become unreasonable.  For instance, it has been demonstrated that fuzzing can effectively discover or recover unknown SCADA commands of a SCADA device \cite{BHS08}.  Physical break-in’s to power substations have been occasionally reported \cite{T14}.  Even worse, obscurity and isolation are no longer effective in the envisioned smart grid, wherein SCADA systems have to coexist and possibly communicate with other relatively more open networks, in order to implement two-way flow of information and energy.  Operation in isolation could be an overly strong assumption for SCADA systems to operate nowadays against the smart grid backdrop.

Exploitation of software vulnerabilities on field device computing platforms is a necessary, key step to compromise field devices in a SCADA system.  This is the only possible way which allows an attacker to install his code on a target machine and spread it to others.  The belief that embedded processors used in field devices have relatively less exploitable software vulnerabilities or it is difficult to compromise an embedded system has become untenable, as demonstrated by the case of the Stuxnet worm discovered in 2010, which has a rootkit to infect PLCs over a proprietary SCADA system and damaged a number of centrifuge equipment in an Iranian nuclear facility \cite{L11,NM18}.  Subsequent generations of Stuxnet were also discovered.  In fact, it has been demonstrated that with crafted instructions and data, compromising an embedded processor is within reach to attackers in general.  These compromises could normally go undetected by the MTU since, as demonstrated by the Stuxnet worm, a compromised device could play man-in-the-middle to conceal all malicious activities from the MTU.

In addition, should a physical break-in to a substation or remote site be possible, an attacker could simply tap in his own attacking devices directly onto the SCADA network to manipulate other innocent devices.  With the increasing connectivity with other systems in the smart grid, this threat is particularly real since these newer smart grid systems typically allow more open access to general users.  For instance, a smart meter may be accessible to a household user’s computer for reading meter data and setting consumption patterns and alerts.  Similarly, an electric vehicle (EV) charging station in a public car park could have a digital interface open for the general public for functionalities like making reservation or checking charging status of an EV.

\section{Data Diode Approach}
\label{sect:data-diode}
A data diode is a unidirectional gateway, similar to a firewall, which sets a digital barrier to enforce network perimeter control for a restricted network (such as a SCADA system) and fend off unwanted accesses from the less secure networks it is connected to (such as smart grid systems).  Yet a data diode provides a stronger security guarantee than a firewall since strict one-way communication --- from the restricted network to the less secure network only --- is enforced by a certain physical law rather than digital logic, with little chance of reverse communication.  Data diodes are often built using fibre optics coupler or transceivers, through the removal of the transmitter component from one side of the communication and the respective receiver component from the opposite side.\footnote{Although different hardware implementations of a data diode exist, supporting different physical channels (e.g. RS-232, EIA-485, USB, Ethernet), most implementations make use of optical couplers to guarantee physical isolation.}  This makes it physically impossible to compromise such devices to achieve reverse connectivity.  Moreover, they usually do not contain firmware, thus requiring minimal or no configuration at all, or have minimal software supported by micro-kernels that can be formally verified.  Whereas, firewalls are often prone to configuration mistakes and relatively accessible for exploit by skilful attackers.  While firmware upgrade is regularly needed in firewall maintenance, patching is seldom needed for data diode deployment.  Finally, data diodes are the only devices receiving the EAL7 (Evaluation Assurance Level 7), the highest grade in the Common Criteria \cite{CC19} (an international standard evaluating the level of security of equipment).

\vskip 1cm
\begin{figure}[htbp]
    \centering
    \includegraphics[width=14cm]{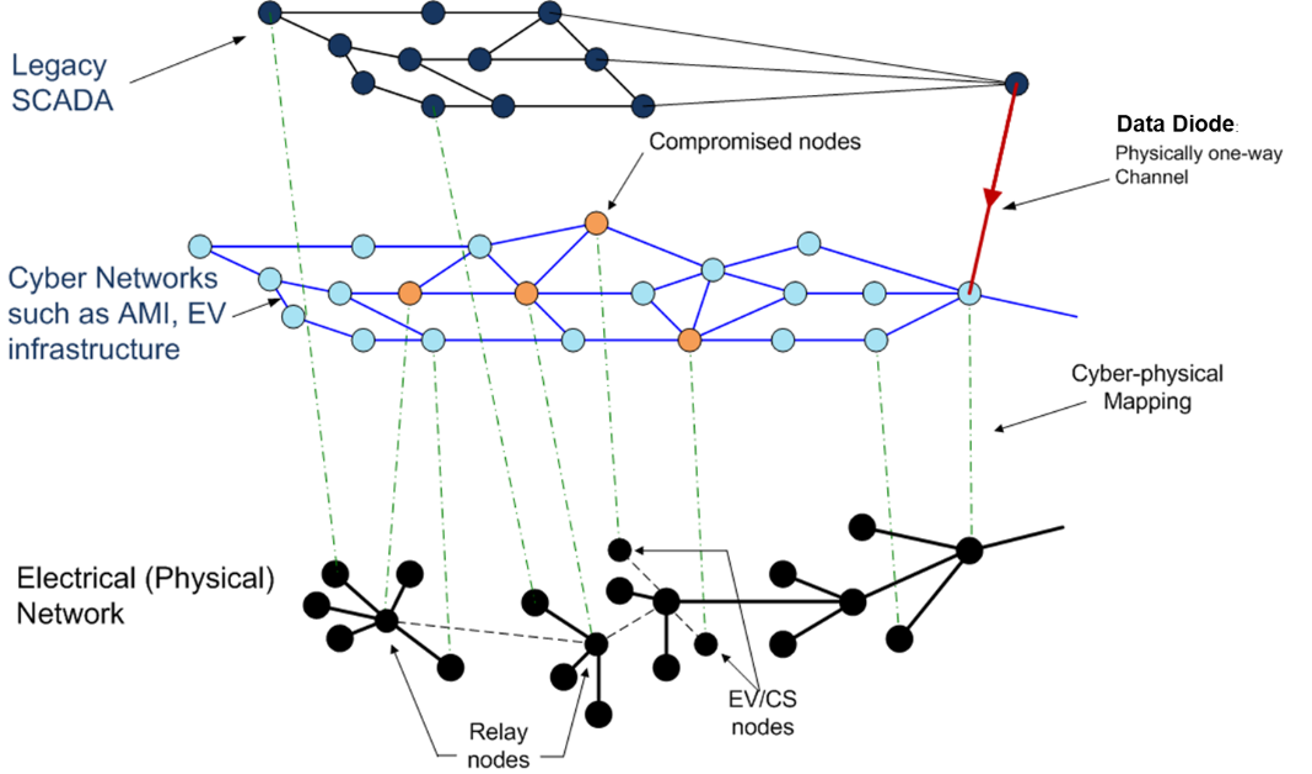}
    \caption{Deployment scenario of the data diode approach}
    \label{fig:data_diode}
\end{figure}

\begin{figure}[htbp]
    \centering
    \includegraphics[width=14cm]{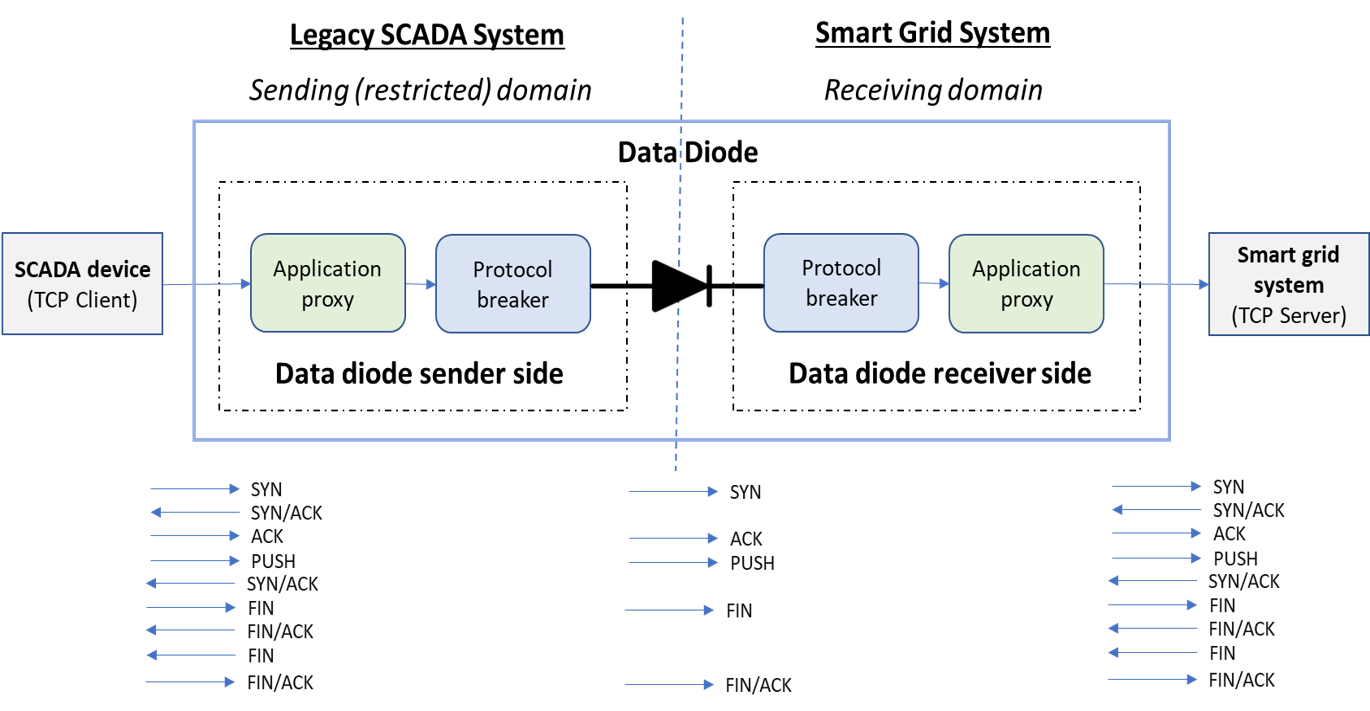}
    \caption{TCP message flow in a unidirectional gateway \cite{FRCS19}}
    \label{fig:data_diode_tcp}
\end{figure}

Data diodes allow organisations to retrieve valuable data generated at the process level from a critical infrastructure for consumption by a wider group of users in a relatively more open system while assuring the isolation and integrity of the critical infrastructure.  As shown in Figure~\ref{fig:data_diode}, a data diode allows data acquired by SCADA field devices to be pushed from the SCADA system so that they can be combined with other datasets captured by smart grid systems like AMI and EV systems which poll the physical conditions of the same electric power grid, in order to give a more holistic and precise picture of the conditions of the electric power grid for more accurate data analysis and hence resource planning.  The correlation of these different datasets can also provide valuable insights for fault identification and predictive maintenance.  Should the analysis lead to desired actions to be carried out in the SCADA system, the feedback will be through human communications.  While existence of compromised nodes is possible in the smart grid systems, the data diode guarantees that they cannot send in false commands to field devices in the legacy SCADA system or infect/compromise them.

Despite similarities in the isolation mechanism, commercially available data diode solutions differ in supported services and protocols.  In order to support TCP/IP-based SCADA protocols, such as MODBUS/TCP and DNP3, data diodes usually have to make use of additional software components for each side of the unidirectional link to emulate the bidirectional message flow for the three-way handshake and acknowledgements needed in a typical TCP session.  As demonstrated by \cite{FRCS19} and shown in Figure~\ref{fig:data_diode_tcp}, the software components include an application proxy and a protocol breaker.  The application proxy emulates the TCP server on the sender side to respond to the SCADA device (TCP client) with TCP connection-oriented messages and acknowledgements while forwarding messages from the TCP client to the unidirectional link.  Similarly, the application proxy emulates the TCP client on the receiver side.  But it does not forward any message from the TCP server on the receiver side to the unidirectional link.  Note that only TCP connection messages sent from the SCADA system to the smart grid system are allowed by the data diode.  The protocol breaker acts as a middleware for packet encapsulation and possibly applying encryption and forward error correction (FEC) to data transferred over the unidirectional link.  FEC is especially important to ensure correct data delivery for the noisy environment of typical SCADA systems.

\section{Detect-and-Respond Approach}
\label{sect:detect-and-respond}
While data diodes provide strong security and isolation/decoupling assurance for legacy SCADA systems, sometimes they are too restrictive for certain application scenarios of the smart grid.  Neither is it optimal to rely on human communications for feedback to a SCADA system.  There is a realistic need for two-way flow of data between a SCADA system and a smart grid system.  Proposals such as \cite{YCKK16} therefore emerged to add a reverse channel to data diodes.  But whether the advantages of a data diode over a typical firewall can be preserved in such a modification is highly uncertain.  More importantly, to what degree that the security assurance of a data diode is undermined with the addition of a reverse channel needs to be assessed carefully.  In scenarios requiring bidirectional data exchange between a SCADA system and a smart grid system, it is therefore more reasonable to assume the existence of compromised devices in the SCADA system \cite{HMW17,HLLL17,LSKSI16,HM18,MKWDSMK16,NPP17,NM18} and devise mechanisms to detect false commands \cite{HMW17,HLLL17,LSKSI16,MKWDSMK16} and/or identify compromised nodes \cite{MKWDSMK16,NPP17}.  In general, a detect-and-respond strategy is preferred despite that many of the proposed schemes in the literature merely cover attack detection.  Once a false command is detected, its impacts should be voided or neutralised promptly, and ideally, in an automatic manner with minimum human intervention required.

This paper presents a generic, high-level architectural framework to implement the detect-and-respond strategy for legacy SCADA systems.  An embodiment in the form of a trusted protection agent is also given to illustrate typical steps needed to neutralise the effects of false commands injected by compromised field devices.  The notion of the protection agent is similar to that of a trust node in \cite{HM18}.  The difference is that the protection agent actively neutralises the effects of false commands, whereas, the trust node only serves as a trusted routing agent to selectively relay messages between field devices and the master station.  Since the framework is defined in a general setting, it should be applicable to different types of SCADA protocols and provide a basis for systematically crafting defence mechanisms fit for different SCADA systems and field devices.  Besides, there is flexibility for incorporating different detection algorithms or mechanisms such as \cite{HMW17,LSKSI16} into the framework.  The framework is also complementary to and could be combined with other approaches \cite{AANL11,ALZR11,CATO17} for securing SCADA systems.

The underlying assumption for implementing the detect-and-respond strategy with a protection agent is that the protection agent is well protected with different protection mechanisms in place to implement the defence-in-depth strategy so as to minimize the probability of its compromise.  In practice, there are various conventional techniques to achieve the security hardening of the protection agent effectively, which will be discussed in Section~\ref{ssect:detect-and-respond:pa} and \ref{ssect:detect-and-respond:security-analysis}.  In the worst case, if a physical break-in to a power substation happens, the protection agent could be compromised.  But the question is, if an attacker successfully breaks in to a power station, he could actually tap in his own device to launch any type of attacks.  The only reason for him to compromise the protection agent is to stop the protection agent from alerting the master station in order to conceal his attacks.  That is, even for physical break-in's, the protection agent would only increase the difficulty for an attacker to launch attacks sneakily.

\subsection{Protection Agent}
\label{ssect:detect-and-respond:pa}

\begin{figure}[htbp]
    \centering
    \includegraphics[width=14.5cm]{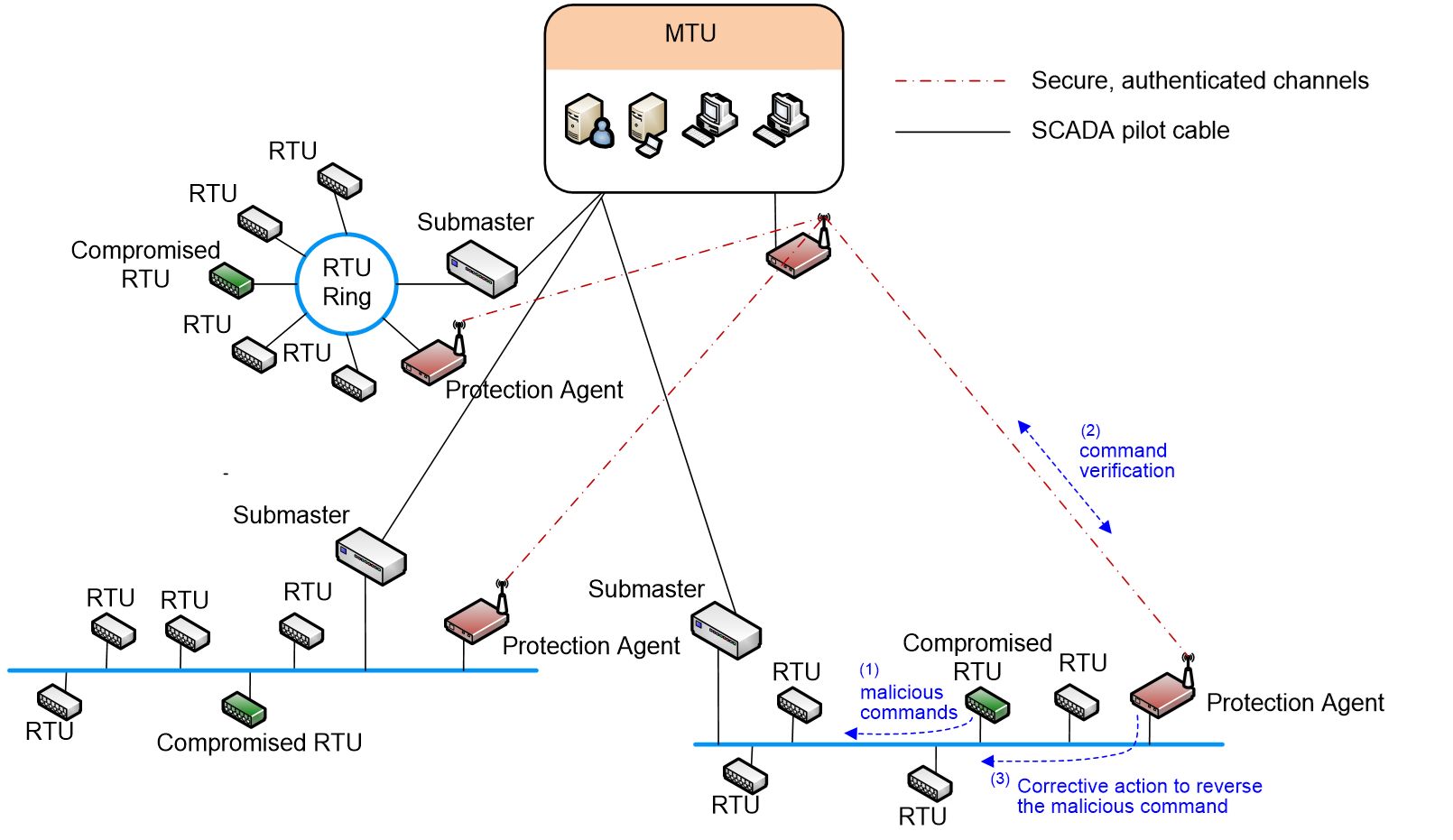}
    \vskip -0.3cm
    \caption{A SCADA system with protection agents installed}
    \label{fig:detect_and_respond}
\end{figure}

As depicted in Figure~\ref{fig:detect_and_respond}, a protection agent is installed as an add-on device in each field network of a SCADA system, which can be made up of PLCs, RTUs or IEDs.  Leveraging on the broadcast medium typically found in most field networks, the protection agent --- as a network sniffer --- monitors commands issued to all the field devices on the network to detect malicious activities.  As a safe assumption, the protection agent is generally a much more powerful machine than typical field devices, given that the underlying processor used in the protection agent could be generations away from those used in the field device of a legacy SCADA system.  In addition, compared to the low-bandwidth physical channel used by these field devices (typically, in the range of 1.2-19.2 kbps \cite{S01,M06,WKM04}), the protection agent can easily be equipped with a wireless link of much higher bandwidth to the MTU or master station.  Note that the design of the protection agent does not preclude other communication channels such as a fiber link.\footnote{A wireless channel is assumed here simply because it is one of the common approaches for adding new communication channels between a remote substation and a control centre.}
We can therefore assume that the protection agent has a faster (with a higher bandwidth and a lower latency) and typically more reliable communication channel (for most of the time) with the MTU than the field devices.  On the basis of a more powerful machine with a faster communication channel to the MTU, security hardening on the protection agent is much easier.  We assume that the protection agent has a secured, authenticated channel with the master station.  Multi-factor entity authentication (for example, those based on pre-shared secret keys stored in a tamper-resistant device, physical unclonable functions \cite{PRTG02,HYKD14}, or even new approaches \cite{CWZT16} etc.) could be adopted to implement this authenticated channel to strengthen its security assurance.  Compared to the field device platforms with little resources for the implementation of cryptographic defence, the protection agent is more ready to implement cryptographic algorithms and protocols, and other security mechanisms requiring more resources (such as firewall or remote code attestation).  In short, the protection agent can be viewed as a trusted proxy for the MTU.

\vskip 1cm

On the other hand, the protection agent can be seen as a reliable incognito for the MTU, monitoring what is happening in the field network.  It is assumed that the protection agent can interpret the underlying protocol messages (as coded in its design) and can sniff messages in the field network.  As discussed in Section~\ref{sect:false_command}, devices in the same field network are usually connected over a broadcast channel, meaning that such traffic sniffing is a practical assumption.  Difference in topology would only affect the number devices that can be eavesdropped.  With the protection agent monitoring the traffic in place, the previously observed problem of some legacy SCADA systems that the MTU is not aware of a circuit breaker having been tripped by an attacker since it has not initiated the tripping \cite{H13} would not happen if a protection agent is installed in the field network.  When the protection agent observes the tripping command in the field network but does not receive the same command through the trusted communication channel shared with the MTU, it would report the tripping event back to the MTU, which would then be aware of a possible intrusion.  While there is possibility that an attacker might physically remove the protection agent from a field network, security mechanisms could be implemented on the protection agent to detect any malicious removal.  For instance, the MTU can run a challenge-response verification protocol with the protection agent by sending a sequence of specially crafted random messages (i.e. nonces) --- destined at a void device identity --- to the protection agent via the field network; if the protection agent is disconnected from the field network, it would not be able to receive these messages to respond to the MTU’s challenge correctly.  Besides, remote code attestation could be implemented to provide the assurance that the software being executed on the protection agent has not been tampered with.  It should be emphasized that a protection agent offers more resources and flexibility than a typical field device like an RTU or PLC for implementing preventive security measures and intrusion detection mechanisms.

\subsection{Detect-and-Respond Defence of Protection Agent}
\label{ssect:detect-and-respond:pa-defence}

\begin{figure}[htbp]
    %\vskip -0.3cm
    \centering
    \includegraphics[width=15cm]{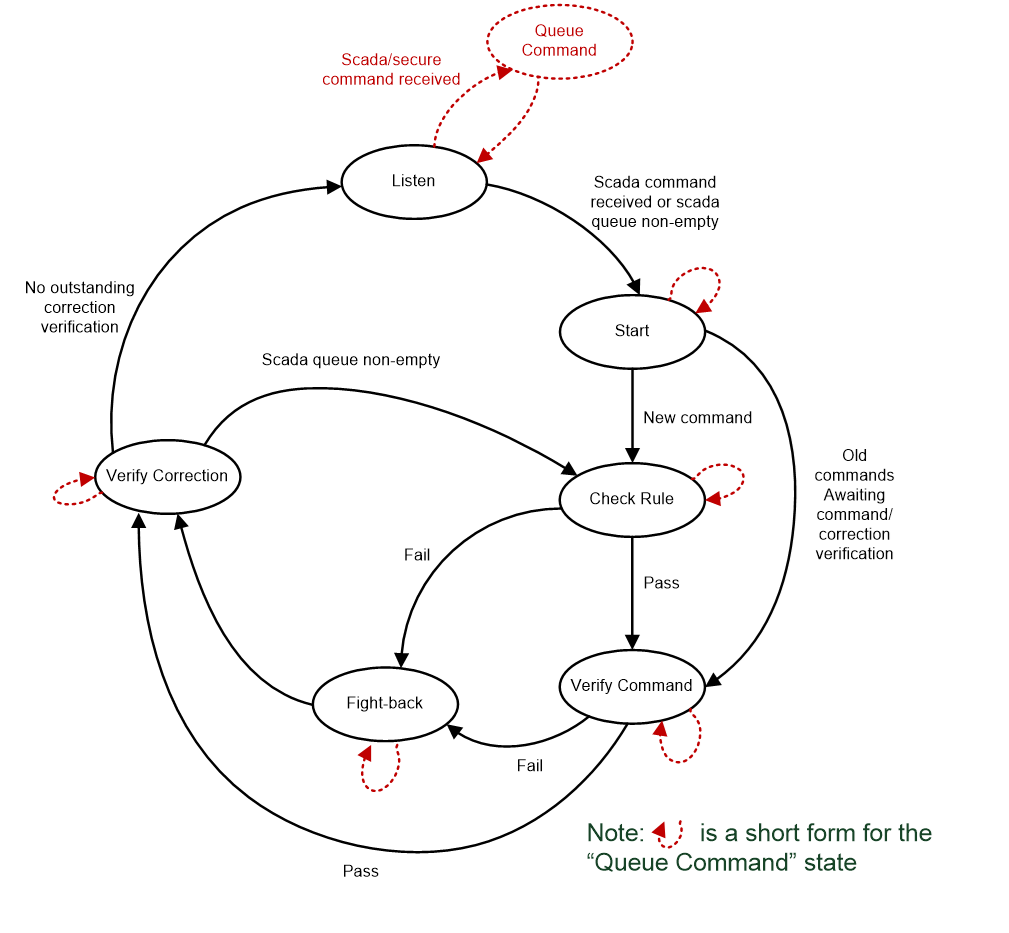}
    \vskip -0.5cm
    \caption{The state machine diagram of a SCADA protection agent}
    \label{fig:PA_state_machine}
\end{figure}

The detect-and-respond mechanism of the protection agent can simply be implemented as a finite state machine as shown in Figure~\ref{fig:PA_state_machine}.  The basic idea is as follows:
\begin{enumerate}
\item Whenever the MTU issues a SCADA command to any field device, it also sends an authenticated message of the same command (protected by a certain cryptographic algorithm such as AES-CCM) to the corresponding protection agent which is connected to the same field network as the concerned field device is in.
\item The protection agent listens or eavesdrops in the field network, sniffing all SCADA commands issued to the field devices in the same field network, and verifies the authenticity and integrity of each of these SCADA commands by comparing them with the authenticated commands directly received from the MTU in the authenticated channel shared between the protection agent and the MTU.
\item For any forged or false command detected (say, by comparing the command received in the field network and that received in the secured channel), the protection agent, possibly after verifying with the MTU, issues a SCADA command to the affected field device to reverse the action caused by the false command to cancel out any undesirable action initiated by the attacker.  This is called the “fight-back” or neutralization mechanism as the protection agent takes action --- in response to a detected malicious command --- to correct it.
\end{enumerate}

\begin{table}[htbp]
\setlength{\tabcolsep}{5pt}
\centering
\begin{tabular}{|p{2.5cm}|p{4.5cm}|p{6.5cm}|}
  \hline
    \textbf{State} & \textbf{Description} & \textbf{Tasks}\\
  \hline

    \emph{Queue} & State to queue newly received SCADA and authenticated commands &
    \begin{minipage}[t]{\linewidth}
        \begin{itemize}[leftmargin=10pt, rightmargin=5pt, itemsep=0pt, parsep=0pt]
            \item read commands from the receiving port
            \item queue the command to the SCADA or authenticated queue accordingly
        \end{itemize}
        \vskip 0.5cm
    \end{minipage}\\
  \hline

  \emph{Listen} & Idle state when the SCADA queue is empty &
  \begin{minipage}[t]{\linewidth}
    \begin{itemize}[leftmargin=10pt, rightmargin=5pt, itemsep=0pt, parsep=0pt]
         \item check if the SCADA queue is non-empty and trigger the “Start” state when the queue is non-empty
    \end{itemize}
    \vskip 0.5cm	
  \end{minipage}\\
  \hline

  \emph{Start} & State to tell between a new SCADA command received or an old SCADA command pending the completion and confirmation of corrective actions taken \vskip 0.5cm &
  \begin{minipage}[t]{\linewidth}
    \begin{itemize}[leftmargin=10pt, rightmargin=5pt, itemsep=0pt, parsep=0pt]
        \item check whether a new SCADA command arrives and trigger command verification if positive
    \end{itemize}
  \end{minipage}\\
  \hline

  \emph{Check Rule} & State to check the local rule set to see if the received SCADA command is illegitimate and violates the prescribed rules &
  \begin{minipage}[t]{\linewidth}
    \begin{itemize}[leftmargin=10pt, rightmargin=5pt, itemsep=0pt, parsep=0pt]
        \item evaluate the SCADA command against the local rule set
        \item if the SCADA does not pass any of the rules, trigger the correction process. Otherwise, check the authenticity of the SCADA command
    \end{itemize}
    \vskip 0.5cm	
  \end{minipage}\\
  \hline

  \emph{Verify Command} & State to check whether the received SCADA command is really from the MTU through the use of the authenticated channel &
  \begin{minipage}[t]{\linewidth}
    \begin{itemize}[leftmargin=10pt, rightmargin=5pt, itemsep=0pt, parsep=0pt]
        \item if the SCADA command has a matched authenticated command received in the authenticated queue, issue a pass, otherwise, request the MTU to verify through the authenticated channel and check again
        \item if the SCADA command fails, trigger the correction process
    \end{itemize}
    \vskip 0.5cm	
  \end{minipage}\\
  \hline

  \emph{Fight-back} & State to correct or reverse the actions caused by a malicious command &
  \begin{minipage}[t]{\linewidth}
    \begin{itemize}[leftmargin=10pt, rightmargin=5pt, itemsep=0pt, parsep=0pt]
        \item look up from a pre-set table for the corrective actions needed for a particular SCADA command
        \item issue the listed SCADA commands to the affected RTUs to reverse the actions of the malicious command
    \end{itemize}
    \vskip 0.5cm	
  \end{minipage}\\
  \hline

  \emph{Verify Correction} & State to confirm that the correction is in effect &
  \begin{minipage}[t]{\linewidth}
    \begin{itemize}[leftmargin=10pt, rightmargin=5pt, itemsep=0pt, parsep=0pt]
        \item read the status of the affected RTUs to confirm that the corrective actions have been applied
    \end{itemize}
    \vskip 0.5cm	
  \end{minipage}\\
  \hline
\end{tabular}
\vskip 0.3cm
\caption{Details of the protection agent implementation as a state machine}
\label{table:pa-implementation}
\end{table}

In principle, the neutralization mechanism could invalidate or void out forged commands, restoring the field device to the original state.  But the degree of neutralization or residual effect of a false command would depend on the actual implementation of the logic and the SCADA command set of a field device.  This defence approach shares some similarities with the fight-back mechanism of OSPF (Open Shortest Path First) --- a link state routing protocol used in the Internet --- through which an innocent router could publicly renounce a malicious route update forged by an attacker \cite{NKGB12}.

In case of missing commands, the protection agent can request the MTU to resend a command over the secured, authenticated channel for verification.  Besides, the implementation of the protection agent could be further extended by including additional false command detection mechanisms beyond comparison with an authenticated copy of a legitimate command.  For example, the detection mechanisms of \cite{HMW17,LSKSI16} can possibly be used by the protection agent to trigger neutralization, preferably after confirmation by the MTU.  However, in order to respond correctly to any maliciously injected commands, the deployed detection mechanism should provide sufficient details to help identify the commands falsely injected by the attacker.

In some cases, a forged or malicious command could be immediately detected in the local field network without relying on authenticated messages from the MTU.  For example, the remote software update of a field device is normally prohibited by most utility operators.  However, in some SCADA systems composed of RTUs, the denial of a remote software update request is implemented on the MTU side only, while the RTUs in the field network usually have no mechanism to deny a software update request or distinguish whether it is initiated by the MTU or an intruding device.  In other words, a compromised RTU or an intruding attacker device could still initiate a remote software update at any RTU in the same field network, which will not be denied by the latter.  This flaw would greatly facilitate the spread of worms or malicious instruction sequences over a field network.  In a typical SCADA deployment, a firewall usually only safeguards a field network from external attacks and would not be able to stop such an insider threat since the attack is launched from within the field network by a compromised field device or an intruding device connected to the field network.  In contrast, the protection agent could possibly halt the malicious software update.  A local rule set can be implemented on the protection agent, explicitly specifying that only manual update is allowed.  When the protection agent detects an automatic software update command, it can immediately halt the update by issuing another command, without having to check with the MTU.

The same technique can be used for other clearly harmful actions.  For instance, the protection agent can enforce that a certain range of parameters is prohibited in some SCADA commands.  With the local rule on protection agents, the damage of the centrifuge equipment caused by the Stuxnet worm could be avoided, since the command with parameters causing excessive spin speed could be invalidated by the protection agent.  Similarly, the protection agent can be configured to eliminate or invalidate set point commands --- which are used in typical SCADA systems to preset the thresholds on certain measured system variables (such as voltages or phase angles at certain point of a power grid) for triggering protective actions like tripping a circuit breaker --- with harmful parameters.

In details, two command queues are implemented in the protection agent, which respectively stores SCADA commands received in the field network and authenticated commands received through the secured, authenticated channel between the MTU and the protection agent.  When the SCADA command queue is empty, the protection agent is at the “Listen” state waiting for new SCADA commands.  The arrival of a new SCADA command will trigger the protection agent to verify the authenticity of the requested action and reverse it in case it is a fake command.  A list of tasks to be implemented at different states of the protection agent is shown in Table~\ref{table:pa-implementation}.

The detect-and-respond strategy deviates from intuition and standard techniques used in securing a communication network.  Whereas existing schemes would suggest filtering all bogus messages injected by an attacker, the proposed mechanism aims to tolerate the intrusion for a while if the resulting malicious actions are not critically harmful and then correct or neutralize the malicious actions.  For critical actions which are usually banned by norm, the local rule set in the protection agent would stop them immediately from onset.  Since the detect-and-respond approach favors local, distributed defence, it is more scalable than centralized defence mechanisms.

\subsection{Example Implementation with Siemens Sinaut 8FW Protocol}
\label{ssect:implementation-on-sinaut}

\begin{table}
  \centering
  \small
  \begin{tabular}{|p{6.5cm}|p{1.4cm}|p{6.5cm}|}
    \hline
    \textbf{Telegram Type} & \textbf{Priority to neutralize} & \textbf{Protection agent's actions to neutralize} \\
    \hline

    \begin{minipage}[t]{\linewidth}
      \begin{itemize}[leftmargin=10pt, rightmargin=5pt, itemsep=0pt, parsep=0pt]
        \item control (Type 64)
        \item replace command (Type 195)
      \end{itemize}
    \end{minipage}& High &
    \begin{minipage}[t]{\linewidth}
      \begin{itemize}[leftmargin=10pt, rightmargin=5pt, itemsep=0pt, parsep=0pt]
        \item Topple the command
        \item Report the action back to MTU and wait for acknowledgement. If negative acknowledgement is received, topple the command back.
      \end{itemize}
      \vskip 0.5cm
    \end{minipage}\\
    \hline

    \begin{minipage}[t]{\linewidth}
      \begin{itemize}[leftmargin=10pt, rightmargin=5pt, itemsep=0pt, parsep=0pt]
        \item analogue/digital set point (Type 65-67/68-70)
        \item modification of threshold value limit (Type 205)
        \item modification of smoothing factor (Type 206)
        \item remote parameterization (Type 212)
      \end{itemize}
    \end{minipage}& High &
    \begin{minipage}[t]{\linewidth}
      \begin{itemize}[leftmargin=10pt, rightmargin=5pt, itemsep=0pt, parsep=0pt]
        \item Temporarily store a copy of the suspected command
        \item Apply the latest confirmed parameter setting stored in the protection agent to form a command
        \item Use replace command (Type 195) to switch the parameter setting back
        \item Report the action back to MTU and wait for acknowledgement.  If negative acknowledgement is received, apply the stored suspected command, otherwise, erase stored command
      \end{itemize}
      \vskip 0.5cm
    \end{minipage} \\
    \hline

    \begin{minipage}[t]{\linewidth}
      \begin{itemize}[leftmargin=10pt, rightmargin=5pt, itemsep=0pt, parsep=0pt]
        \item start-up request (start up/restart) (Type 211)
        \item switch on/off recipient in master station (Type 203)
        \item switch off record transfer from/to station (Type 204)
      \end{itemize}
      \vskip 0.5cm
    \end{minipage}& Medium &
    \begin{minipage}[t]{\linewidth}
      \begin{itemize}[leftmargin=10pt, rightmargin=5pt, itemsep=0pt, parsep=0pt]
        \item Report to the MTU and wait for confirmation.  If confirmed, topple the command.
      \end{itemize}
      \vskip 0.5cm
    \end{minipage}\\
    \hline

    \begin{minipage}[t]{\linewidth}
      \begin{itemize}[leftmargin=10pt, rightmargin=5pt, itemsep=0pt, parsep=0pt]
        \item switch on/off for temporal lists (Type 201-202)
        \item synchronization of fine time (type 207)
        \item setting of minutes (Type 208)
        \item setting of calendar (Type 209)
        \item switch on/off addresses in the lists (Type 210)
        \item 4-byte-storage interrogation control (Type 214)
        \item interrogation command ZFBIT and STOP-cause (Type 215-222)
      \end{itemize}
      \vskip 0.5cm
    \end{minipage}& Low &
    \begin{minipage}[t]{\linewidth}
      \begin{itemize}[leftmargin=10pt, rightmargin=5pt, itemsep=0pt, parsep=0pt]
        \item Report to the MTU and wait for confirmation.  If confirmed, topple the command or restore the stored parameters.
      \end{itemize}
    \end{minipage}\\
    \hline

    \begin{minipage}[t]{\linewidth}
      \begin{itemize}[leftmargin=10pt, rightmargin=5pt, itemsep=0pt, parsep=0pt]
        \item check command (Type 192)
        \item check command (type 192)
        \item message repeat request / TFK-acknowledge (Type 193)
        \item start acknowledgement (Type 194)
        \item single/group interrogation command (Type 196-197)
        \item multiple request (Type 198-200)
        \item matrix-check command (Type 213)
      \end{itemize}
      \vskip 0.5cm
    \end{minipage}& Low &
    \begin{minipage}[t]{\linewidth}
      \begin{itemize}[leftmargin=10pt, rightmargin=5pt, itemsep=0pt, parsep=0pt]
        \item Neutralization is not necessary.
      \end{itemize}
    \end{minipage}\\
    \hline

  \end{tabular}
  \normalsize
  \vskip 0.3cm
  \caption{Implementation of the detect-and-respond defence on Siemens Sinaut 8FW protocol}
  \label{table:sinaut_embodiment}
\end{table}

Table~\ref{table:sinaut_embodiment} illustrates how the detect-and-respond defence strategy can be implemented with respect to the command set of the Siemens Sinaut 8FW protocol \cite{S01}, a popular industrial protocol commonly used in legacy SCADA systems.  The Sinaut 8FW protocol was the standard protocol used by Siemens devices and systems for communicating control and monitoring messages between MTU systems and RTU devices before being replaced by IEC 60870-5-101.  It was broadly supported by devices provided by other manufacturers like ABB and GE.  While the principle of the detect-and-respond strategy can be equally applied to newer protocols like IEC 60870-5-101 and 60870-5-104, its implementation with these protocols is outside the scope of this paper, partly because it is possible to implement authentication mechanisms in these protocol \cite{PKK13}.  In contrast, the resource constraints of devices running the Sinaut 8FW protocol preclude the possibility of implementing any security mechanism despite that the protocol is still actively used in a non-negligible fraction of systems.  In other words, it is more critical to investigate add-on protection for the Sinaut 8FW protocol over others.
Each command is enclosed in a control message in the form of a telegram.  Table~\ref{table:sinaut_embodiment} only shows telegram types sent in the control direction (from the MTU to field network), with other telegram types in the monitoring direction (from the field network to the MTU) skipped since these telegram types are largely for messages sent to the MTU.

False commands that may have critical impact on the functioning of a SCADA system are neutralized first and then verified with the MTU, whereas, the less critical ones are corrected upon confirmation from the MTU.  Most of the neutralization actions can be readily implemented through the replace command telegram.  For commands which switch on/off a certain feature or function, the neutralization can simply be done by toppling the command in the oppositive direction with respect to the false command.  For commands which involve updating of parameters like thresholds or set points to trigger preventive actions, the protection agent stores the latest version of all the parameters of a field device (with parameters of different field devices stored in different tables) and apply relevant ones to form a replace command when necessary to revert the parameters affected by a false command.

Experimentation was carried out on an Intel 2.5GHz Quad Core Celeron CPU (with 8GB RAM) --- which simulates the protection agent --- to estimate the response time of the protection agent to issue a neutralization command or telegram for different types of false commands.  The protection agent is connected to a server (which simulates the master station) through a WiFi link.  On the other side, the protection agent is connected to another single board computer through a RS232 serial port (set at a baud rate of 19200), which simulates a compromised field device (in Scenario 1) or a master station (in Scenario 2) to issue commands to the protection agent.  Two types of response time are measured corresponding to two different scenarios.  In the first scenario (i.e. Scenario 1), the protection agent is set to receive no message from the master station and the compromised field device (simulated by the single board computer) injects a false command to the serial port to trigger the protection agent to neutralize the command.  The response time needed to issue a neutralization command after the injection of a false command is measured.  In the second scenario (i.e. Scenario 2), the single board computer simulates an authentic command from the master station.  That is, the injected command is authentic but the protection agent mistakenly considers it as a false command.  The protection agent first issues a neutralization command and then queries the master station which always returns a negative acknowledgement.  That is, the protection agent has to switch back the command and cancel the effects of the neutralization command.  The response time for the protection agent to issue the second command after the injection of the first command by the single board computer is measured.

\begin{figure}[htbp]
    \centering
    \includegraphics[width=14cm]{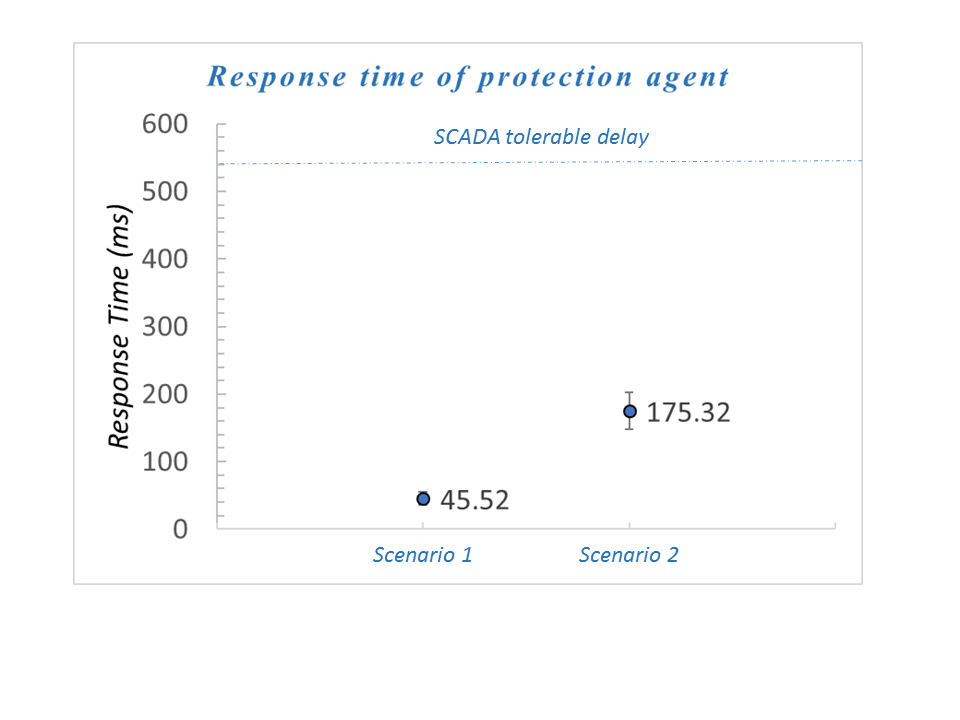}
    \vskip -1.5cm
    \caption{Response time of the protection agent for: (a) Scenario 1: Neutralization of a malicious command; (b) Scenario 2: Ratification of false neutralization.}
    \label{fig:response_time}
\end{figure}

The experimental measurements are presented in Figure~\ref{fig:response_time}.  The average response time required for the protection agent to look up the parameter for a particular field device identity and issue a neutralization command or telegram for different types of false commands (at high priority) is 45.52 ms with a standard deviation of 9.28 ms.  The average response time required for the protection agent to switch back the command for the second scenario is 175.32 ms with a standard deviation of 27.5 ms.
Compared to the maximum tolerable delay for SCADA transactions, which is 0.54s \cite{BBT05}, this processing delay is acceptable (respectively, 8.4\% and 32.4\% of the maximum tolerable delay).

\subsection{Security Analysis}
\label{ssect:detect-and-respond:security-analysis}

The key assumption for the detect-and-respond mechanism based on a protection agent is that the protection agent is trusted and has a reliable communication channel to the master station.  This should be a reasonable assumption in practice.

First, it is considerably harder to compromise a protection agent than a typical field device.  But it is easier to detect any compromise of the protection agent.  As discussed in Section~\ref{ssect:detect-and-respond:pa}, since the protection agent is based on a more powerful computing platform compared to a field device, various techniques --- such as remote code attestation to ensure that a correct version of software is executed on the protection agent and challenge-response verification to ensure that the protection agent is not disconnected from the field network --- can be readily applied to strengthen the security assurance of the protection agent and detect any compromise.  Besides, if a protection agent is compromised, the attacker can cause no more harm than what he can achieve with a compromised field device when no protection agent is deployed.  Since compromising a protection agent is more difficult than compromising a field device, adding a protection agent would make the job of an attacker more difficult and strengthen system security.

Second, the more powerful platform used for a protection agent (compared to a field device) enables the implementation of common cryptographic primitives, thereby allowing the use of standard protocols such as the TLS (Transport Layer Security) to secure the communication channel between a protection agent and the master station.  The confidentiality and authenticity of each message sent over the channel is thus assured.
An out-of-band wireless channel is assumed for the communication between a protection agent and the master station because it is the most cost-effective way, and a popular means, to add a new communication channel to a power substation.  To address the reliability issues associated with a wireless channel, the protection agent can request the master station to resend any missed command or explicitly acknowledge/confirm any command it has sent through the field network.  Since such a channel usually has a higher bandwidth and a lower latency compared to the field network, the processing delay is usually reasonable.  As demonstrated by the experimental results, the latency for the confirmation required for a missed command (Scenario~2) seems practically acceptable.  Alternatively, an in-band, authenticated communication channel \cite{CATO17} or a more reliable communication channel such as a fiber link could be used.

\subsection{Cost and Benefit Analysis}
\label{ssect:detect-and-respond:advantages}
The detect-and-respond strategy embodied by the protection agent offers a number of advantages as follows.
\begin{enumerate}
\item The protection mechanism is non-intrusive and truly add-on while strong cryptographic mechanisms can be adopted to protect the communications between a field network and the MTU (via a protection agent), without replacing any field devices or modifying their code.  No laying of new cables is needed, as wireless links like mobile cellular networks could be used between protection agents and the MTU.
\item Compared to the bump-in-the-wire approach [1, 14] which adds a new cryptographic device per field device, a smaller number of protection agents are required, with one required for each field network in most cases.  A typical field network could have at least 30 devices, whereas, the maximum number of field devices per network is limited by the protocol's address space.  As an example, up to 128 and 65535 devices per network are allowed for Sinaut 8FW and IEC60870-5-104 respectively.  Hence, compared with the bump-in-the-wire approach, the detect-and-respond approach can reduce the number of cryptographic devices by a factor between 1/128 and 1/30.  The minimum number of protection agents needed for a SCADA system is roughly equal to the number of its field networks.  For instance, less than 100 protection agents are needed to secure the SCADA system of Singapore's 22kV distribution grid (with $\sim10,000$ substations and the IEDs connected in multiple field networks in a ring topology), whereas, 1-2 protection agents suffice to cover that of its 66kV distribution grid (with $\sim100$ substations and the field devices connected through a mesh topology).  In other words, there are about 100 field networks (with each in the ring topology) for the SCADA system of the 22kV distribution grid, and only one field network in the mesh topology for the SCADA system of the 66kV distribution grid.
\item The protection agent would not become a bottleneck since additional protection agents can be deployed to a given field network with many field devices (say, more than 300 devices).  In such cases, the set of field devices in a network can be grouped into partitions based on their identities and multiple protection agents could be installed in the network with each responsible for a distinct partition.  Since any number of protection agents can be added to a field network and the field devices need not be aware of the presence of a protection agent, the solution is scalable.
\item The detect-and-respond strategy is particularly effective for defending against insider attacks, which existing solutions like firewalls often cannot withhold.  A firewall normally cannot filter messages sent from a field device located behind it or forged messages from outside if no authentication mechanism is in place. Unlike a firewall, the protection agent does not present a single point of failure.  If it is down, the field network would still operate as usual, which is not the case for a firewall.
\end{enumerate}

\section{Conclusions}
\label{sect:conclusions}

The instrumentation and control of a power grid is usually carried out by a SCADA system which is a distributed network of cyber-physical devices taking measurements and issuing commands at different parts of a power grid.  Obscurity and operation in isolation have long been the security strategy of SCADA systems.  However, this is no longer a reasonable assumption for SCADA systems in the smart grid context with the need of connectivity with other relatively open networks.  False command injection is a real threat to legacy SCADA systems.  While effective security mechanisms and algorithms for command authentication are largely an overkill for implementation on field devices in legacy SCADA systems, these devices would normally take decades to be phased out as they are closely integrated with the power equipment.  This paper discusses and compares two different approaches, namely, the data diode strategy and the detect-and-respond strategy, to secure legacy SCADA systems for safe integration with other smart grid systems.  A detailed comparison of the two approaches can be found in Table~\ref{table:comparison}.

\begin{table}
\centering
\begin{tabular}{|p{2.5cm}|p{4.6cm}|p{4.6cm}|}

\hline
                         & \textbf{Data diode} & \textbf{Detect-and-respond} \\
\hline
    Security assurance   &  Highest (EAL7) & Medium to High \\
\hline
    Applicable cases & Limited range & Wide range\\
\hline
    Pros &
    \begin{minipage}[t]{\linewidth}
        \begin{itemize}[leftmargin=13pt, rightmargin=5pt, itemsep=0pt, parsep=0pt]
            \item Isolation preserved with a high level of security
            \item No modifications on field devices required
            \item Highly scalable, with only one data diode required per system
        \end{itemize}
    \end{minipage} &
    \begin{minipage}[t]{\linewidth}
        \begin{itemize}[leftmargin=13pt, rightmargin=5pt, itemsep=0pt, parsep=0pt]
        \setlist{nolistsep}
           \item Flexible to incorporate other techniques for attack detection and neutralization
           \item Allow bidirectional data exchange
           \item No modification on field devices required
           \item Scalable
        \end{itemize}
        \vskip 0.3cm
    \end{minipage}\\
\hline
    Cons &
    \begin{minipage}[t]{\linewidth}
        \begin{itemize}[leftmargin=13pt, rightmargin=5pt, itemsep=0pt,parsep=0pt]
         \item Only allow unidirectional data flow from SCADA systems to other systems
         \item Little flexibility
       \end{itemize}
    \end{minipage} &
    \begin{minipage}[t]{\linewidth}
        \begin{itemize}[leftmargin=13pt, rightmargin=5pt, itemsep=0pt, parsep=0pt]
         \item More complex design required to maintain security
         \item Design strongly dependent on the SCADA protocols in use
       \end{itemize}
       \vskip 0.3cm
    \end{minipage}\\
\hline
\end{tabular}
\vskip 0.3cm
\caption{Comparison between the data diode and detect-and-respond strategy}
\label{table:comparison}
\end{table}

The detect-and-respond strategy proposed in this paper presents a number of advantages for securing legacy SCADA systems, including scalability and usability for a wide range of smart grid scenarios.  The proposed framework is also generally applicable to different SCADA protocols and systems, while a concrete instantiation on Siemens Sinaut 8FW protocol is presented in this paper.
While the data diode strategy offers the highest level of security guarantee, it can only be applied to a restricted number of use cases wherein unidirectional data exchange suffices.  In contrast, the detect-and-respond strategy is a more flexible approach which can be applied in almost all scenarios but generally offer a lower level of security guarantee and is more complex to design.

%\nocite{*}
\bibliographystyle{plain}
\bibliography{./scada}

\end{document}